\documentclass{aa}  

\usepackage{graphicx}
\usepackage{xcolor}
\usepackage{txfonts}
\usepackage{subcaption}         
\usepackage{lscape}             
\usepackage{placeins}           
                              
\usepackage[colorlinks=true, linkcolor=blue, citecolor=blue, urlcolor=blue]{hyperref}
\usepackage{lipsum}

\newcommand{\be}{ \begin{equation} }
\newcommand{\ee}{\end{equation}}

\def\PGPU{$\varphi-$GPU }

\begin{document}

\title{The contribution to Galactic Centre $\gamma$-ray excess from cluster-born millisecond pulsars}

\subtitle{Constraints from direct N-body simulations}


   \author{M. Kalambay\inst{1,2,3} 
        \and M. Ishchenko\inst{4,5,2,1}
        \and D. Kuvatova\inst{2,3}
        \and T. Panamarev\inst{6,2,1}\thanks{Corresponding author: \email{panamarevt@gmail.com}}
        \and P. Berczik\inst{5,2,7,4}
        }

    \institute{
    Heriot-Watt University Aktobe Campus, K. Zhubanov Aktobe Regional University, A. Moldagulova ave. 34, 030000 Aktobe, Kazakhstan \email{\href{mailto:M.Kalambay@hw.ac.uk}{M.Kalambay@hw.ac.uk}}
    \and
    Fesenkov Astrophysical Institute, Observatory 23, 050020 Almaty, Kazakhstan
    \and  
    Faculty of Physics and Technology, Al-Farabi Kazakh National University, al-Farabi ave. 71, 050040 Almaty, Kazakhstan
    \and 
    Main Astronomical Observatory, National Academy of Sciences of Ukraine, 27 Akademika Zabolotnoho St, 03143 Kyiv, Ukraine
    \and
    Nicolaus Copernicus Astronomical Centre, Polish Academy of Sciences, ul. Bartycka 18, 00-716 Warsaw, Poland
    \and 
    Rudolf Peierls Centre for Theoretical Physics, Parks Road, OX1 3PU, Oxford, UK
    \and
    Szechenyi Istvan University, Space Technology and Space Law Research Center, H-9026 Gyor, Egyetem ter 1. Hungary
    }

\titlerunning{$\gamma$-ray excess from cluster-born millisecond pulsars.}

\authorrunning{Kalambay et al.}

\date{Received September 30, 20XX}

\abstract
{The Galactic-Centre $\gamma$-ray excess (GCE) is a spatially extended surplus of $\gamma$-rays around Sgr A* observed by \textit{Fermi}-LAT that exceeds predictions from standard cosmic-ray interactions in the inner Galaxy. The GCE has been mainly attributed either to dark-matter annihilation or to an unresolved population of millisecond pulsars (MSPs) dynamically delivered to the inner Galaxy by globular clusters (GCs).}
{We reassess the MSP interpretation by following a fully dynamical framework to explore how neutron stars (NSs) produced in GCs are stripped and deposited into the central kiloparsec. We also quantify the resulting $\gamma$-ray emission to compare it with \textit{Fermi}-LAT measurements.}
{We ran high-resolution direct $N$-body simulations of GCs in a time-varying Milky Way (MW) potential, capturing the internal cluster dynamics and external tides. We modelled two channels: (i) present-day observed MW GCs on their orbits and (ii) a population of early, now-destroyed clusters whose debris was accreted during their evolution into the inner Galaxy. The simulations provide the full phase-space distribution of deposited NSs to central 1 kpc from these both sources. We converted the NS count to an MSP count using an empirically calibrated efficiency, adopting a sample-averaged ratio between observed and predicted MSP counts and  NSs that are bound to clusters. As a result, we were able to generate mock sky maps and cumulative flux profiles assuming representative per-pulsar luminosities.}
{We find that MSPs inferred from the observed GCs already supply a substantial $\gamma$-ray signal, including disrupted clusters increases both amplitude and central concentration. By taking the observational ratio of MSPs to the total number of NSs in observed Galactic stellar systems as a basis and adopting an average luminosity of $\langle L\rangle\!\sim\!8\times10^{33}$\,erg\,s$^{-1}$, we were able to reproduce the GCE quite well. As a free parameter to achieve better agreement with the observed flux, we had to increase the number of NSs originating from previously disrupted stellar systems by approximately a factor of 2.
The deposited NSs from destroyed clusters exhibit an axisymmetric morphology with pronounced over-densities in the Galactic plane and perpendicular to it.}
{The results of our modelling favour an MSP origin of the GalC $\gamma$-ray excess over dark matter annihilation, primarily because the combined contribution of MSPs delivered by surviving and disrupted GCs naturally reproduces both the amplitude and concentration of the observed signal under reasonable assumptions.}


   \maketitle

\section{Introduction}

The Galactic Centre (GalC) is a region of intense astrophysical activity, hosting the supermassive black hole (SMBH) Sagittarius A* \citep{Ghez2008, Gillessen2017}, a dense nuclear stellar cluster \citep{Neumayer2020}, complex interstellar gas and radiation fields, and intense cosmic-ray activity, as reviewed in \citealt{Genzel2010}. It is expected to be the brightest source of diffuse $\gamma$-ray emission in the sky, primarily arising from interactions of high-energy cosmic rays with the interstellar medium (ISM) and radiation fields. However, observations have revealed an anomalous surplus of $\gamma$-ray emission in the GeV energy range that exceeds predictions from conventional astrophysical models. This signal, known as the GalC GeV Excess (GCE), was first identified in 2009 using the initial year of data from the \textit{Fermi} $\gamma$-ray Space Telescope's Large Area Telescope (\textit{Fermi}-LAT), an instrument with wide field-of-view, high angular resolution, and energy coverage from $\approx$~100 MeV to >300 GeV \citep{Goodenough2009,Hooper2011}. The excess has since been robustly confirmed by independent analyses, including those from the \textit{Fermi}-LAT Collaboration, using over a decade of accumulated data, establishing it as a spatially extended, spectrally peaked feature accounting for ~2\% of the total $\gamma$-ray flux within a 30$^\circ$ radius of the GalC \citep{Ackermann2017,DiMauro2021}. The signal peaks at a few GeV, is approximately spherically symmetric about Sgr~A* and is detectable out to $\gtrsim10$--$15^\circ$ (a few kpc) from the centre \citep{Goodenough2009,Hooper2011,Hooper2018,DiMauro2021,Cholis2022}.

The origin of the GCE remains a subject of intense debate, with interpretations being broadly divided between exotic physics and astrophysical sources. One leading hypothesis attributes the excess to the annihilation of weakly interacting massive particles (WIMPs), a candidate for dark matter (DM), concentrated in the GalC due to the cuspy density profile of the Milky Way's (MW) DM halo. This model naturally explains the GCE's bump-like spectrum (peaking at 1 -- 4 GeV), spherical morphology (consistent with a Navarro-Frenk-White profile with inner slope $\gamma \approx 1.2$), and intensity, which aligns with the thermal relic annihilation cross-section of $\langle\sigma v\rangle \approx 10^{-26}$ cm$^3$ s$^{-1}$ \citep{Hooper2011,Daylan2016,DiMauro2021,Cholis2022}. Alternatively, a more conservative explanation posits that the GCE arises from a large, unresolved population of millisecond pulsars (MSPs) in the Galactic bulge, whose collective $\gamma$-ray emission mimics the observed spectrum due to a power-law with exponential cut-off at a few GeV, matching the average spectral properties of known \textit{Fermi}-LAT detected MSPs \citep{Bartels2016,Lee2016,Gautam2022,Holst2025}. \cite{Eckner2018} showed that the GCE is not unique to the MW: a similar extended $\gamma$-ray emission is also observed in M31. The authors demonstrated that a population of MSPs forming in situ in bulges, with their luminosity scaled by stellar mass, can reproduce both the energetics and the morphology of the observed signal.

A prominent astrophysical route to a bulge population of MSPs invokes the long-term dynamical evolution and disruption of globular clusters (GCs). Early works have shown that a large fraction of the MW’s GC system is destroyed over Hubble time, depositing stars and compact remnants into the bulge \citep{Tremaine1975, Gnedin1997, Gnedin2014}. Building on this, \citet{Brandt2015} argued that MSPs delivered by disrupted clusters can quantitatively account for the GCE’s amplitude, spectrum, and approximately spherical morphology. 

Direct $N$-body models that follow infalling clusters with internal mass segregation and tidal stripping support this picture, linking the high-energy emission to compact objects brought in by cluster inspiral and dissolution \citep{Arca-Sedda2014, ArcaSedda2018}. Independent constraints on the MSP yield of GCs, from population analyses of cluster pulsars \citep{Turk2013} to recent FAST discoveries of multiple new MSPs in a single GC, underscore that dense clusters can host substantial MSP reservoirs \citep{Turk2013, Yin2024}. \cite{Fragione2018} performed direct $N$-body simulations (\texttt{NBODY7} + \texttt{SSE} + \texttt{BSE}) of GCs evolution with varying initial masses, metallicities, and primordial binary fractions, tracing the formation and spatial distribution of their neutron stars (NSs) populations over 10 Gyr. They showed that most NSs escape from their parent clusters due to the natal kicks received during Type II supernova explosions, as the escape velocity of the clusters is much lower than the typical kick velocity. Assuming 10\% of the NSs are recycled into MSPs, the authors demonstrated that the resulting $\gamma$-ray luminosity of the GCs matches Fermi observations, supporting the MSP origin of the GCE. Taken together, these results motivate a fresh, fully dynamical assessment of whether GC-borne NSs can indeed reproduce the detailed properties of the GCE.

In our previous work \citep{Kuvatova2024}, we analysed the dynamical evolution of six GCs (Palomar 6, HP 1, NGC 6401, NGC 6642, NGC 6681, and NGC 6981) and estimated the potential contribution of their putative MSPs to the $\gamma$-ray excess in the GalC. We found that even when considering a small number of clusters, their MSPs can provide a noticeable contribution to the excess, while not fully accounting for it, which is the motivation behind the present study.

In this work, we revisit the MSP interpretation by running high-resolution, direct $N$-body simulations of GCs evolving in a time-varying MW potential, explicitly tracing the orbits of all NSs formed in (and stripped from) the clusters. We consider two complementary channels: (i) present-day Galactic GCs on their observed orbits, to test whether repeated pericentre passages and tidal stripping can inject NSs into the central few kiloparsecs where they might be recycled into MSPs and contribute to the GeV signal; and (ii) a population of now-destroyed clusters whose debris was accreted into the central regions, depositing NSs that subsequently power the GCE. Our approach self-consistently captures internal cluster dynamics (e.g. mass segregation) and external tides, and it produces the full phase-space distribution of deposited NSs. We then map these NS populations to the $\gamma$-ray emission using empirically calibrated MSP efficiencies and luminosity functions, and confront the resulting surface-brightness profiles and spectra with \textit{Fermi}-LAT measurements. 

The paper is organised as follows. In Sect. \ref{sec:ini-discr} we describe integration procedure and initial conditions for GCs. Section \ref{sec:ns-galc} presents the NS distribution in Galaxy at the present day based on GCs $N$-body simulations. In Sect. \ref{sec:gamma-flux}, we provide our main results about $\gamma$-ray fluxes from the distribution of MSPs based on NSs from the simulations. Section \ref{sec:disc-con} contains our conclusions. 

\section{Integration procedure, initial mass function, and mass loss}\label{sec:ini-discr}

To investigate the spatial distribution of GCs' NSs around the GalC, we performed detailed full $N$-body simulations with stellar evolution using the high-order parallel code $\varphi$-GPU\footnote{$N$-body code \PGPU: \\~\url{ https://github.com/berczik/phi-GPU-mole}} \citep{Berczik2011, BSW2013}. This code implements a fourth-order Hermite integration scheme with hierarchical individual block time steps and supports parallel computation via CUDA library. In its current version, the code incorporates an updated stellar evolution library \citep{Banerjee2020, Kamlah2022MNRAS}, which accounts for key evolutionary processes, including stellar mass loss via winds, supernova mechanisms, natal kicks, and more. 

Each of our present day clusters was initialised in a state of dynamical equilibrium using a King model \citep{King1966}, characterised by the total initial mass, $M$, the dimensionless central potential, $W_0$, and the half-mass radius, $r_{\rm hm}$. These parameters were numerically adjusted to fit the current mass and size of the clusters. The stellar mass distribution followed the Kroupa initial mass function \citep{Kroupa2001} in the range 0.08--100 M$_\odot$. A more detailed description of the integration procedure and the mass evolution of the investigated GCs can be found in \cite{Ishchenko2024massloss}.

\begin{table}[tbp]
\centering
\caption{Initial physical characteristics at -8 Gyr for MW and early theoretical GCs.}
\label{tab:init-param}
\begin{tabular}{cccccc}
\hline
\hline 
No & GC & M$_{\rm ini}$ & N & r$_{\rm hm}$ & $W_0$ \\
& & $10^{6}\rm\;M_{\odot}$ & & pc & \\
\hline
\hline
1 & NGC 362     & 0.6 &   993 265 & 1.5 & 3.0 \\
2 & NGC 1904    & 1.0 & 1 759 800 & 7.0 & 8.0 \\
3 & NGC~6401    & 1.2 & 2 249 640 & 6.5 & 9.0 \\
4 & NGC~6642    & 1.5 & 2 613 856 & 4.0 & 9.0 \\
5 & NGC~6681    & 1.3 & 2 265 380 & 3.0 & 8.0 \\
6 & NGC~6981    & 1.0 & 1 742 560 & 7.0 & 9.0 \\
7 & NGC 7078    & 0.9 & 1 603 040 & 2.0 & 3.0\\
8 & HP 1        & 1.3 & 2 265 340 & 6.0 & 8.0  \\
9 & Palomar 6   & 1.0 & 1 751 970 & 3.5 & 9.0 \\
10 & Terzan 2   & 1.1 & 1 916 827 & 5.0 & 9.0 \\
11 & Terzan 4   & 1.3 & 2 265 341 & 5.0 & 9.0 \\
12 & Terzan 5   & 2.3 & 4 007 900 & 2.0 & 3.0 \\
\hline 
13 & Destr. GC  & 0.06 & 104 554 & 4.0 & 8.0 \\
14 & Destr. GC  & 0.18 & 313 662 & 2.0 & 8.0 \\
\hline 
\end{tabular}
\end{table}

For the simulations, we selected 12 real GCs. These clusters were chosen primary because of their close passages near the GalC. The initial physical parameters of GCs used in the simulation are listed in Table \ref{tab:init-param}. The initial kinematic conditions for the cluster centres were derived via individual backward orbital integration of GCs as point mass objects to a lookback time of 8 Gyr from the present state, as presented in \cite{Ishchenko2023a}. As external potential in both types simulations we used a MW-like time-variable galactic potential from the IllustrisTNG-100 simulation \citep{Nelson2019, Ishchenko2023a}. 

\begin{figure}[htb!]
\centering
\includegraphics[width=0.99\linewidth]{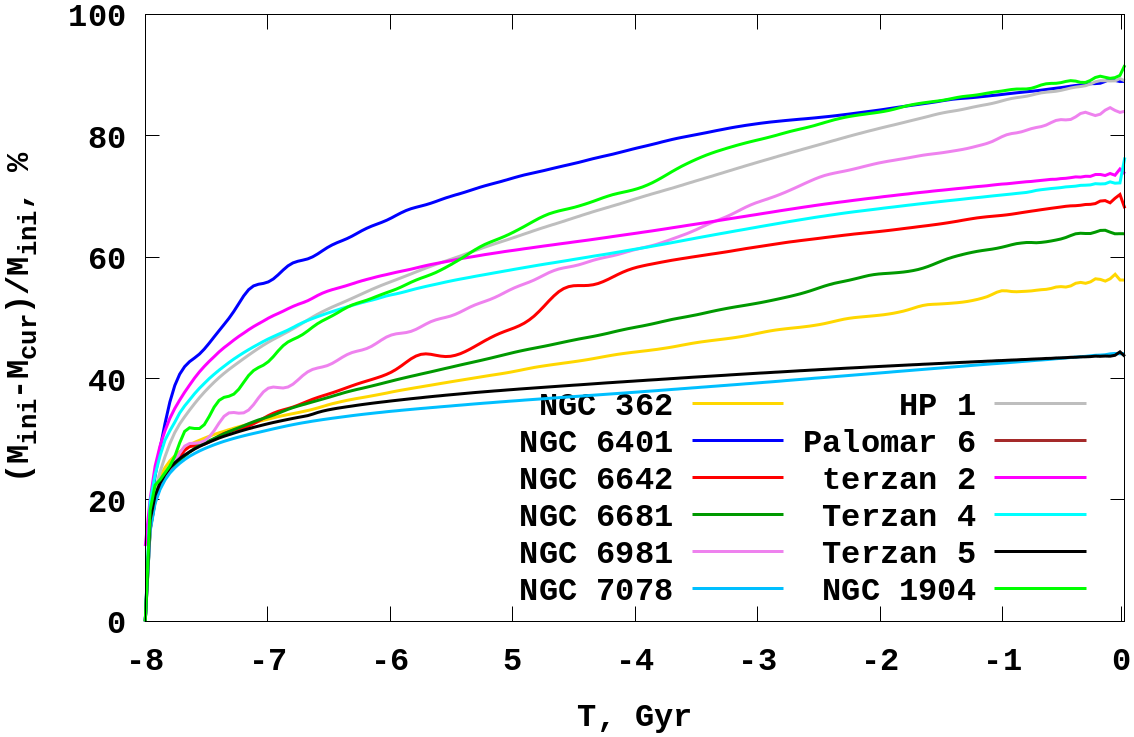}
\caption{Evolution of the mass loss in the GCs in percent due to stellar evolution and orbital tidal stripping for our 12 MW GCs.}
\label{fig:mass-loss}
\end{figure}

Figure \ref{fig:mass-loss} shows the fraction of lost mass for the real GCs over the 8 Gyr of orbital integration. As seen in the plot, the clusters experience the most rapid mass loss during the first billion years of their evolution. The most prominent mass loss is observed for NGC 1904, NGC 6401, HP 1, and NGC 6981.

In addition to the real clusters, we also simulated a set of theoretical GCs. This sample included 50 clusters with an initial mass of 60 000 M$_\odot$ and 4 more massive clusters with a mass of 180 000 M$_\odot$ (see other parameters at the end of Table \ref{tab:init-param}). All of these theoretical clusters were fully disrupted during the 5 Gyr simulation, contributing their stellar content to the MW's central region. In one theoretical GC with a mass of 60 000 M$_\odot$, a total of more than 700 NS are formed, and in a more massive one, three times more. More information about initial conditions and modelling process can be found in \cite{Kuvatova2026}.

Here, we briefly describe the initial positions and velocities of each cluster's centre. For generating the mesh of random initial conditions, we use the global specific binding energy and the two components of the specific angular momentum, $L_{\rm tot}$ and $L_z$, for each disrupted GC. For this purpose, we integrate back 10~Gyr in our well-tested {\tt 411321} Illustris TNG-TVP potential. Using this external potential, we compute the phase-space distribution of the GC centres. We adopt specific binding energies in the range $(-19$ to $-14)\times 10^{4}\,{\rm km^2\,s^{-2}}$. For the specific angular-momentum space, we chose $L_{\rm tot}=0$--$3\times 10^{2}\,{\rm kpc\,km\,s^{-1}}$ and $L_z=(-0.5$ to $+0.5)\times 10^{2}\,{\rm kpc\,km\,s^{-1}}$, spanning both prograde and retrograde orbits and selected to cover the range of currently observed (and modelled) Galactic GCs \citep{Ishchenko2023b, Ishchenko2024massloss}.

\section{NS distribution in Galaxy at the present day based on GC simulations}\label{sec:ns-galc}

The number of NSs in a GC depends both on its initial mass, range of the IMF, and the individual metallicity. In our simulations, the fraction of NSs formed over the course of the cluster’s evolution ranges from 0.5 -- 0.8\% relative to the initial number of stars. At the beginning of GC evolution, the largest number of NSs from the 12 MW GCs with their evolution modelled in Terzan 5 is over 22k. The lowest number of the NSs evolved in NGC 362 is around 5k.   

However, due to the high natal kick velocity prescription, which is implemented in our \textit{N}-body code, fewer than one-fifth of them remain bound to their parent clusters \citep[e.g.][]{Lyne1994, Hansen1997, Podsiadlowski2004, Banerjee2020, Kamlah2022MNRAS}. Recent studies continue to explore and support diverse natal kick prescriptions \citep{Roberts2025, Popov2025}. These recent prescriptions have included  various non-radial asymmetries in supernova explosions, along with very rapid rotation as well as the influence of magnetic field. As a result (taking in account all factors), the retained NS fraction is only about 0.01--0.1\% of the initial stellar population (see more details in Table~\ref{tab:ns-stat}), which is also consistent with previous estimates \citep{Pfahl2002, Kuranov2006}.

We classified the NS as a cluster member if it has a negative binding energy with the nucleus of the cluster (i.e. NSs are bound with the GC). We computed the amount of the NS within a tidal radius. The tidal radius (or `Jacobi' radius) was calculated based on the numerical iteration of the $M_{\rm tid}$ and $r_{\rm tid}$ values using the equation
\be
r_{\rm tid} = \left[ \frac{G \cdot M_{\rm tid}}{M_{\rm Gal}} \right]^{1/3}, 
\ee
where $G$, $M_{\rm tid}$, and $M_{\rm Gal}$ are respectively the gravitational constant, the cluster tidal mass and the Galaxy enclosed mass at the GC current position. For more details, we refer to \cite{King1962, Just2009, Ernst2011}. 

\begin{figure}[htb!]
\centering
\includegraphics[width=0.99\linewidth]{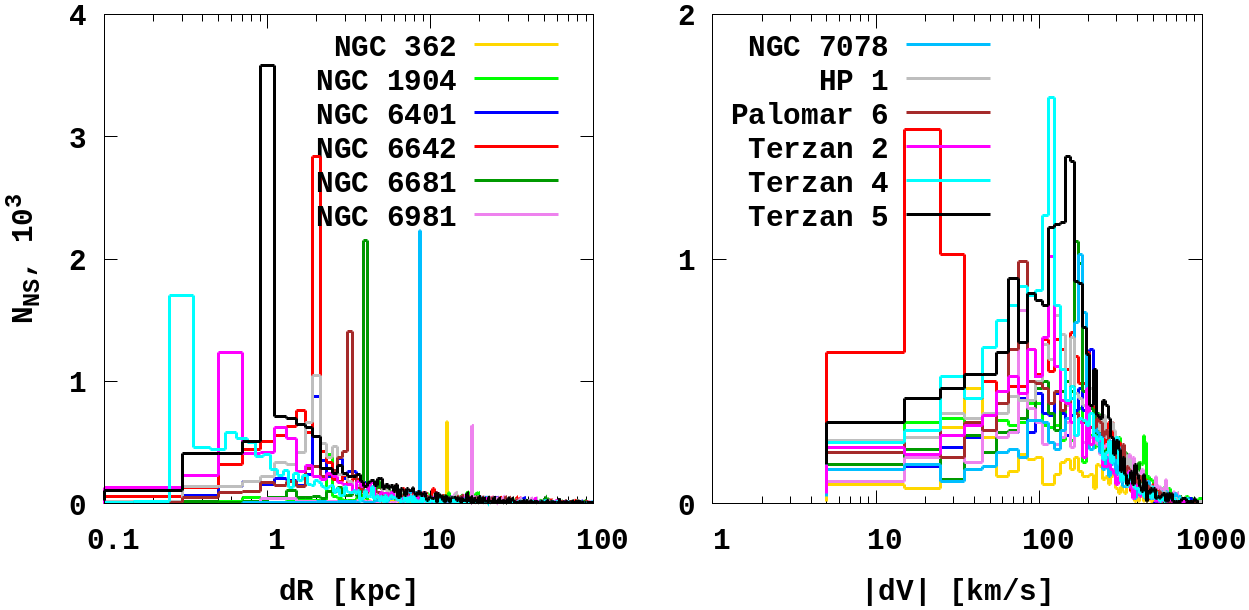}
\includegraphics[width=0.99\linewidth]{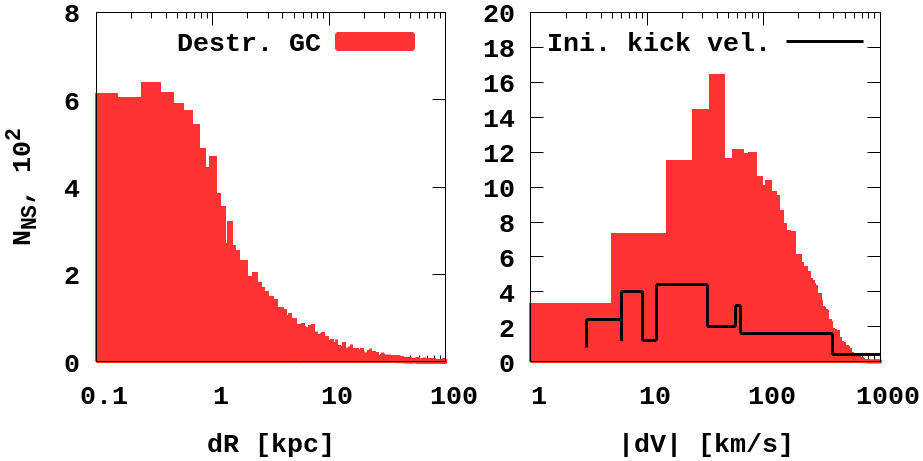}
\caption{Position and velocity distribution of the NSs in the Galaxy from the GC's $N$-body simulations. Top: Real MW GCs. Bottom: Early GCs that were destroyed over a 5 Gyr evolution time.}
\label{fig:dr-gc}
\end{figure}

In Fig.~\ref{fig:dr-gc}, we show the distribution of NSs in the Galaxy and the corresponding velocity values. The top panel shows peaks corresponding to bound NSs within the GC's $r_{\rm tid}$ boundaries at the present day. As expected, the majority of NSs in the GC are found in Terzan 5, NGC 7078, and NGC 6642. The mean relative distance between the GalC and our set of GCs varies from 0.1--10 kpc, with a velocity distribution of 10--500 km s$^{-1}$. We note that the NGC 6642 (red colour) has the lowest velocity values and Terzan 5 has the highest.

In bottom panel of Fig. \ref{fig:dr-gc}, we can see that most of all NS from early destroyed clusters are located in the central region of the Galaxy, less than 1 kpc from the centre. As mentioned above, Sect. \ref{sec:ini-discr}, we modelled 50 GCs that were initially located in the central region of the Galaxy (at less than 5 kpc) and most of them were located even less than 1 kpc from the centre. 

\begin{figure}[htb!]
\centering
\includegraphics[width=0.99\linewidth]{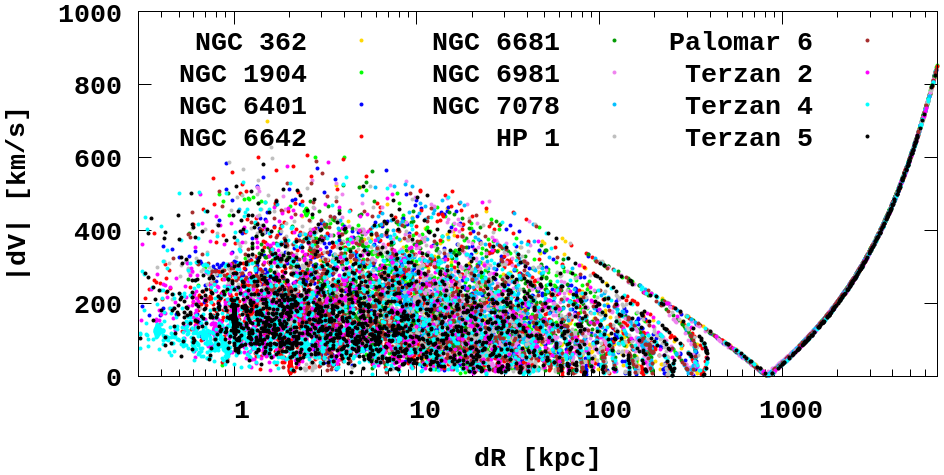}
\includegraphics[width=0.99\linewidth]{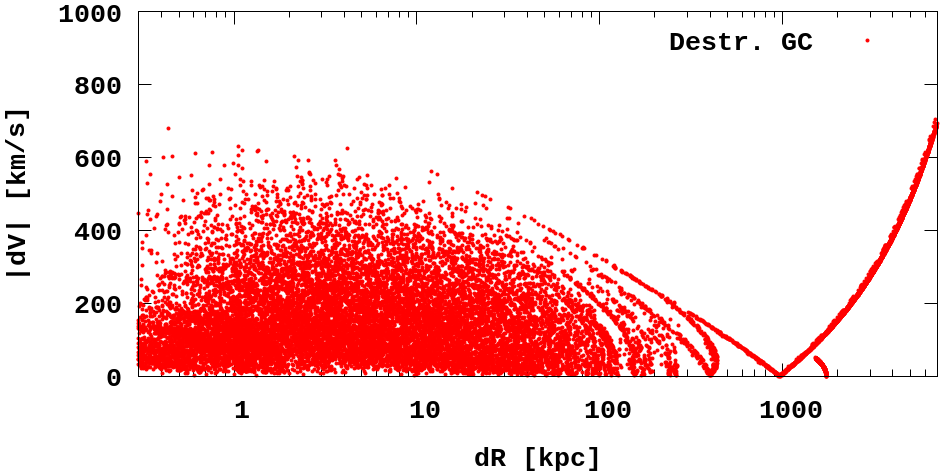}
\caption{Dependence of the velocity distribution on distance. Top: Real MW GCs. Bottom: Destroyed GCs. The colour coding corresponds to the legend in Fig. \ref{fig:mass-loss}.}
\label{fig:ns-distr}
\end{figure}

The dependence of velocity as a function of distance is demonstrated in Fig. \ref{fig:ns-distr}. A small fraction of NSs (1--2\% of all NSs) were generated in 12 MW GCs and became unbound at an early stage of the GCs evolution. And after 8 Gyr of evolution, their relative distance exceeded 1000 kpc and even more. These stars had an initial kick velocity of up to 1000 km s$^{-1}$ at the beginning, as shown by the black line in the bottom panel of Fig. \ref{fig:dr-gc}. As a result, such stars have the potential to completely escape from our Galaxy.

\begin{figure*}[htb!]
\centering
\includegraphics[width=0.95\linewidth]{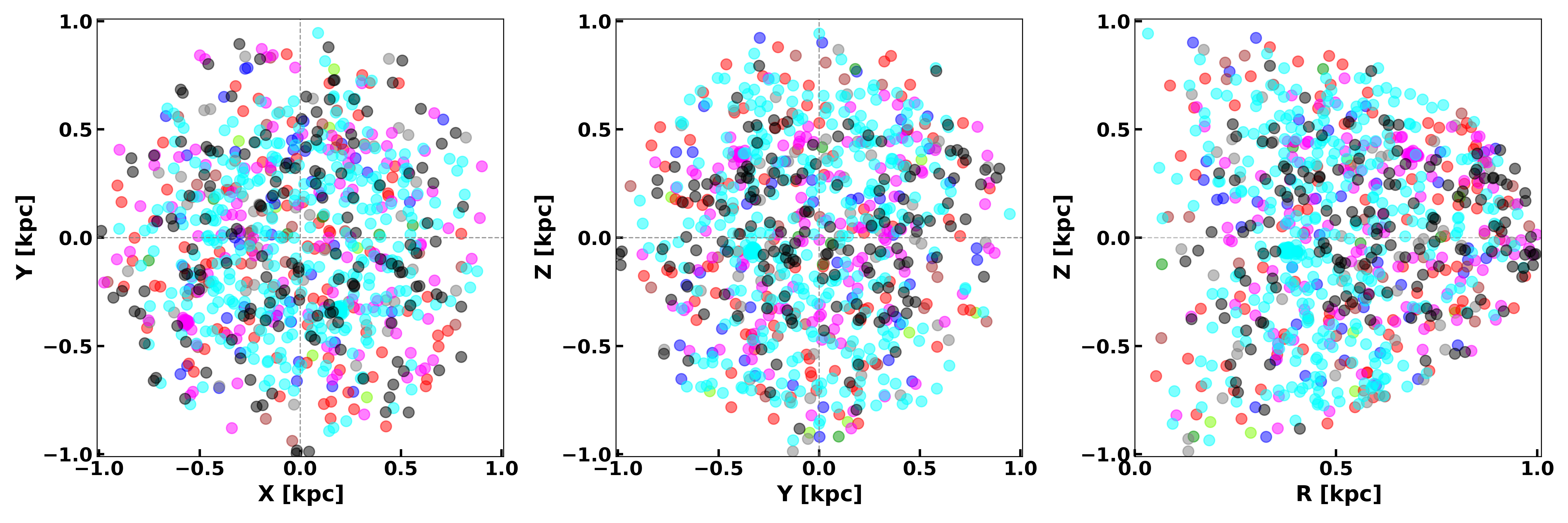}
\includegraphics[width=0.95\linewidth]{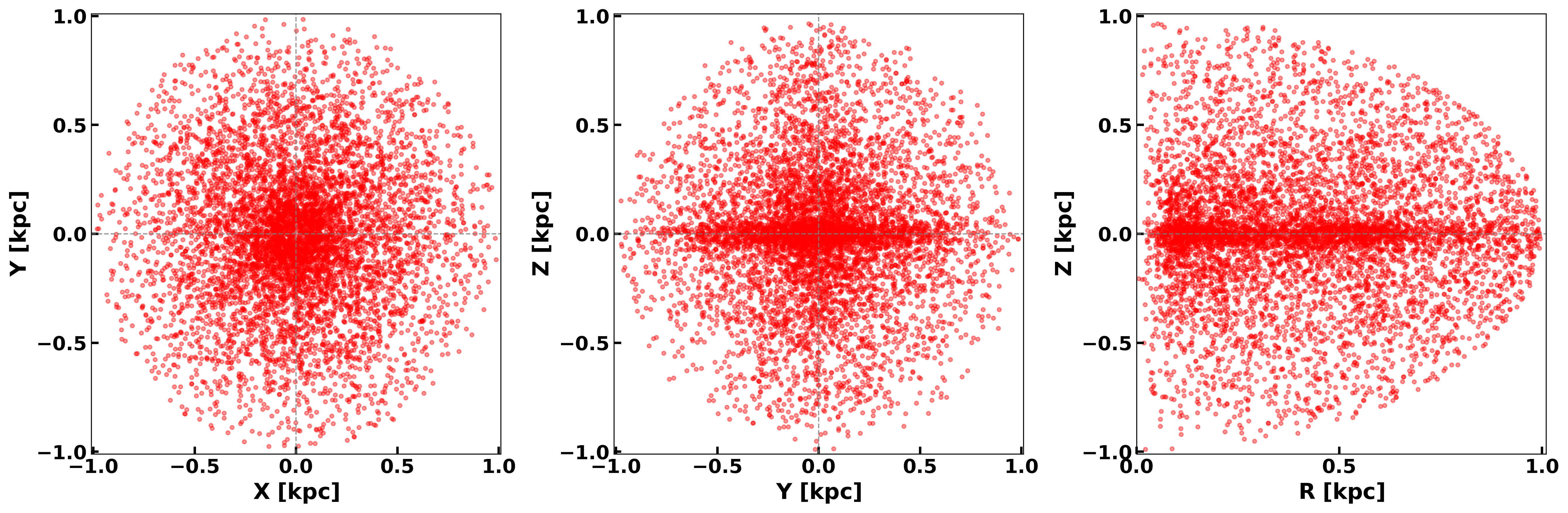}
\caption{Distribution of NSs in the central region of the Galaxy is less than 1 kpc. This distribution is shown in a three-coordinate projection. From left to right: $X$ vs $Y$, $Y$ vs $Z$ and $R$ vs $Z$, where $R$ is the galactic plate radius. Top: NSs from MW GC simulations (different colours correspond to different models, as in Fig. \ref{fig:mass-loss}). Bottom: NSs from GCs that were destroyed early on.}
\label{fig:XYZ-GCs-Real-Destr}
\end{figure*}

In Table \ref{tab:ns-stat}, we summarise all statistical information about the investigated NS. Here, we present the ratio of the  initial NS mass to the mass of bound NSs in GCs after 8 Gyr of orbital evolution (i.e. today). In column 8, we also present the number of observed MSPs in GCs from the web-based online catalogue\footnote{Observed MSPs: \\~\url{ https://www3.mpifr-bonn.mpg.de/staff/pfreire/GCpsr.html}}. We also present the theoretical numbers of MSPs for each MW GC based on the study of \cite{Turk2013} and \cite{Yin2024}. In the last column, we give the numbers of NSs in the central area of the GalC (see also Fig.~\ref{fig:XYZ-GCs-Real-Destr}). 

\begin{table*}[htbp]
\caption{Overview of the NS distribution in MW GCs and early theoretical GCs along with the theoretical number of MSPs.}
\centering
\scriptsize
\resizebox{\textwidth}{!}{%
\begin{tabular}{lcccccccc}
\hline
\hline
GC & Total NS & Ini. M & Ini. NS/M & NS in GC & Cur. M & Cur. NS/M & MSP & NS in box  \\
& & $10^{6}\rm\;M_{\odot}$ & $10^{6}\rm\;M_{\odot}$ &  & $10^{6}\rm\;M_{\odot}$ & $10^{6}\rm\;M_{\odot}$ & obs, (theor1, theor2) & 1000, 500, 250, 150 pc\\
\hline
NGC 362    & 5\,550  & 0.6    & 0.0092 & 690   & 0.252   & 0.0027 & 12 (25, 28)   & 0, 0, 0, 0 \\
NGC 1904   & 13\,550 & 1.0    & 0.0135 & 390   & 0.640    & 0.0006 & 1 (7, 16)     & 70, 0, 0, 0 \\
NGC 6401   & 12\,480 & 1.2    & 0.0104 & 650   & 0.121   & 0.0053 & -- (4, 11)       & 450, 100, 0, 0 \\
NGC 6642   & 17\,730 & 1.5    & 0.0118 & 2\,630& 0.040  & 0.0666 & -- (7, 8)        & 950, 120, 0, 0 \\
NGC 6681   & 13\,340 & 1.3    & 0.0103 & 2\,630& 0.011  & 0.0025 & 3 (31, 13)    & 60, 10, 10, 10, 10 \\
NGC 6981   & 11\,710 & 1.0    & 0.0117 & 980   & 0.008 & 0.0012 & -- (1, 4)        &  0, 0, 0, 0 \\
NGC 7078   & 11\,090 & 0.9    & 0.0123 & 2\,270& 0.518   & 0.0044 & 15 (80, 102)  & 0, 0, 0, 0 \\
HP 1       & 15\,140 & 1.3    & 0.0116 & 520   & 0.137   & 0.0038 & -- (0, 8)        & 490, 70, 10, 10 \\
Palomar 6  & 13\,540 & 1.0    & 0.0135 & 180   & 0.008 & 0.0138 & -- (2, 8)        & 270, 40, 20, 10 \\
Terzan 2   & 12\,030 & 1.1    & 0.0109 & 840   & 0.008 & 0.0104 & -- (3, 7)        &  2010, 230, 0, 0 \\
Terzan 4   & 17\,070 & 1.3    & 0.0131 & 1\,460& 0.181   & 0.0081 & -- (--, 11)       &  4630, 2190, 10, 10 \\
Terzan 5   & 22\,630 & 2.3    & 0.0098 & 2\,930& 1.090    & 0.0027 & 49 (104, 215) & 1470, 300, 30, 0 \\
\hline
1 (50) Destr. GC & 723 (36\,150) & 0.06e6  & 0.0120    & --    & --        & --     & --              & 5211, 2578, 723, 424 \\
\hline
1 (4) Destr. GC & 2153 (8609) & 0.18e6  & 0.0119     & --    & --        & --     & --              & 1711, 725, 152, 95 \\
\hline
\end{tabular}%
}
\tablefoot{Column 8 gives the number of observed MSPs in GCs \footnote{Observed MSPs: \\~\url{ https://www3.mpifr-bonn.mpg.de/staff/pfreire/GCpsr.html}}. In brackets: theoretically predicted number of MSPs according to \cite{Turk2013} and \cite{Yin2024}; see their Tables 4 and 5.}
\label{tab:ns-stat}
\end{table*}

\section{$\gamma$-ray fluxes from the distribution of MSPs}\label{sec:gamma-flux}
\subsection{Fitting the MSP-to-NS ratio}\label{subsec:ns-fitt}

From our cluster simulations, we were able to directly obtain the number of NSs located in the central region of the Galaxy within a maximum radius up to 1 kpc (see Table~\ref{tab:ns-stat}, column~9).

The central NS population is a mixture of stars stripped from different GCs at different epochs. Some of these NSs may end up as MSPs \citep{Bhattacharya1991,Lorimer2008} with the capacity to emit $\gamma$~radiation. Our simulations provide the full 6D phase space for all these NSs; however, they do not include binary stellar evolution, which is required to recycle NSs into pulsars. Nevertheless, we can statistically estimate the number of pulsars per NS (based on the observed MSPs in real GCs) to assess the contribution of these MSPs to the observed $\gamma$-ray excess.

\begin{figure}[htb!]
\centering
\includegraphics[width=0.95\linewidth]{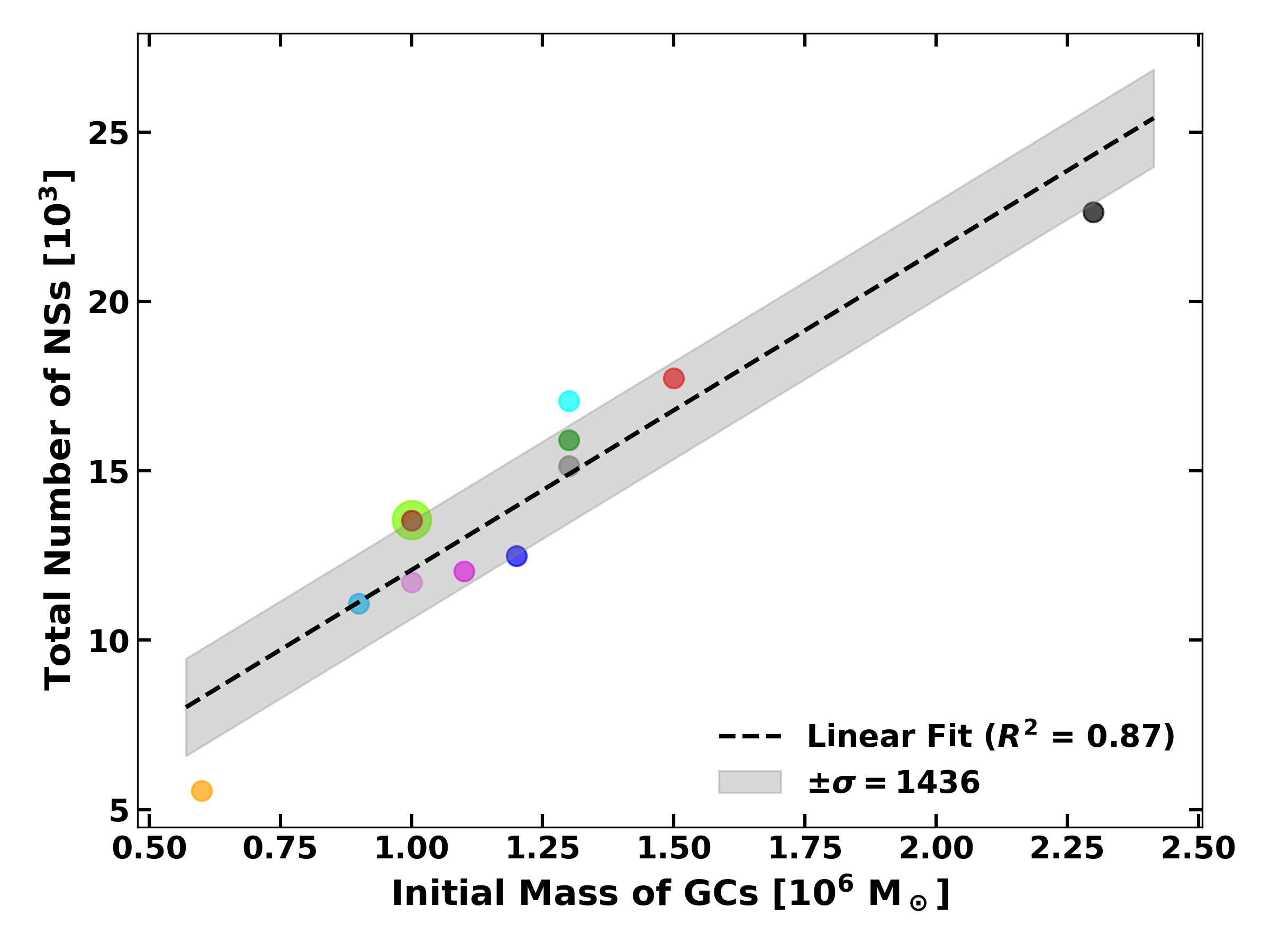}
\includegraphics[width=0.95\linewidth]{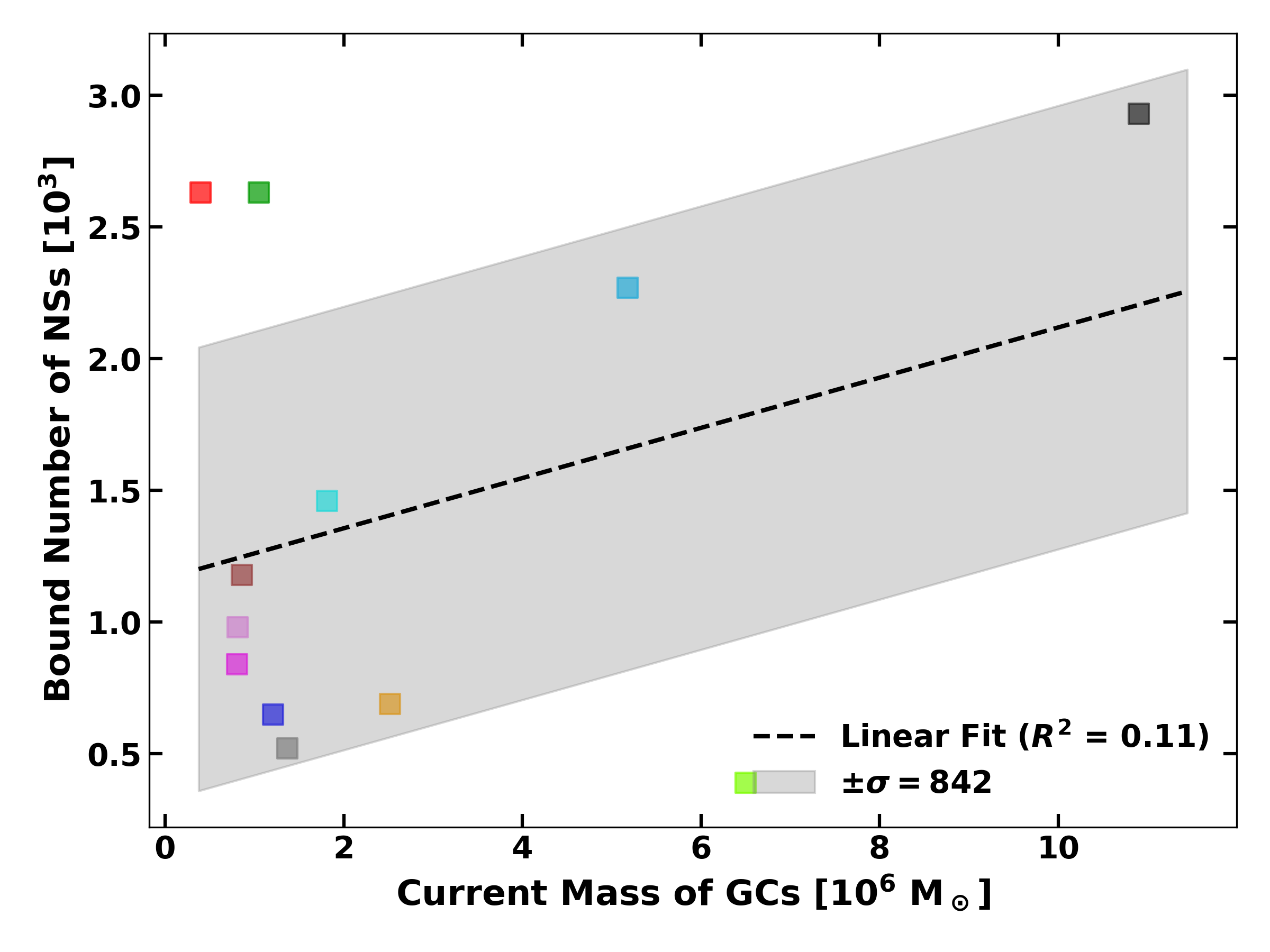}
\caption{Relationship between GC mass and NS populations. Colours denote individual clusters as listed in legend of Fig.~\ref{fig:dr-gc}. The full dataset is given in Table~\ref{tab:ns-stat}. Top: Initial mass of GCs vs the total number of NSs formed up to the present day. The dashed line shows the best-fit linear relation (slope = 9428.42 NSs per $10^6 M_\odot$, $R^2 = 0.87$), with the shaded region representing the $\pm$1436 NSs scatter around the fit. Bottom: Current mass of GCs vs bound number of NSs. The dashed line shows the best-fit linear relation (slope = 95.29 NSs per $10^5 M_\odot$,  $R^2 = 0.11$), with the shaded region representing the $\pm$842 NSs scatter around the fit.}
\label{fig:ratio-GCs-NS}
\end{figure}

A cluster’s initial mass largely sets the total number of NSs formed, but this proportionality strictly applies at formation. Figure~\ref{fig:ratio-GCs-NS} (top) shows that the total number of NSs correlated with the initial cluster mass. Because all our modelled clusters share the same Kroupa initial mass function and stellar-mass range, the total number of NSs produced in a GC is nearly proportional to its initial mass; any differences between clusters of the same initial mass arise mainly from metallicity. By contrast, the correlation weakens when we examine the number of bound NSs as a function of a cluster’s current mass within the tidal radius (Fig.~\ref{fig:ratio-GCs-NS}, bottom panel), reflecting the effects of mass loss on different Galactic orbits, stellar stripping, tides, and retention physics.

To infer the fraction of NSs in the GalC that are observable today as MSPs, we adopted an observationally motivated statistical calibration. Specifically, we mapped the number of bound NSs in each modelled cluster to the number of observed MSPs in the corresponding GC \citep{Turk2013,Yin2024}. Because the observed counts can be incomplete, we also used the predicted total MSP numbers for those clusters \citep{Turk2013,Yin2024}, which both mitigated any selection biases and allowed for the inclusion of clusters without confirmed MSP detections.

\begin{figure}[htb!]
\centering
\includegraphics[width=0.95\linewidth]{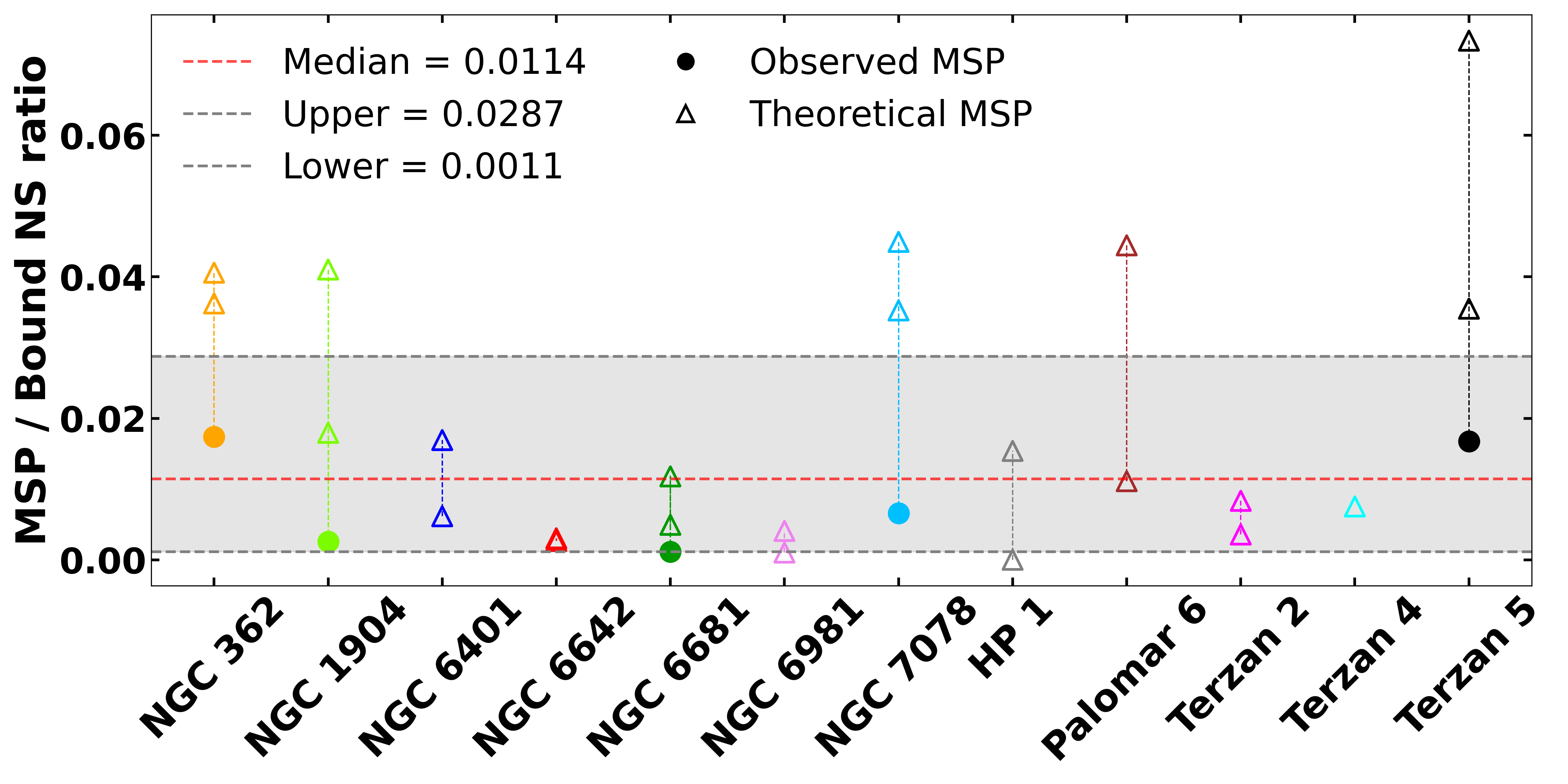}
\includegraphics[width=0.95\linewidth]{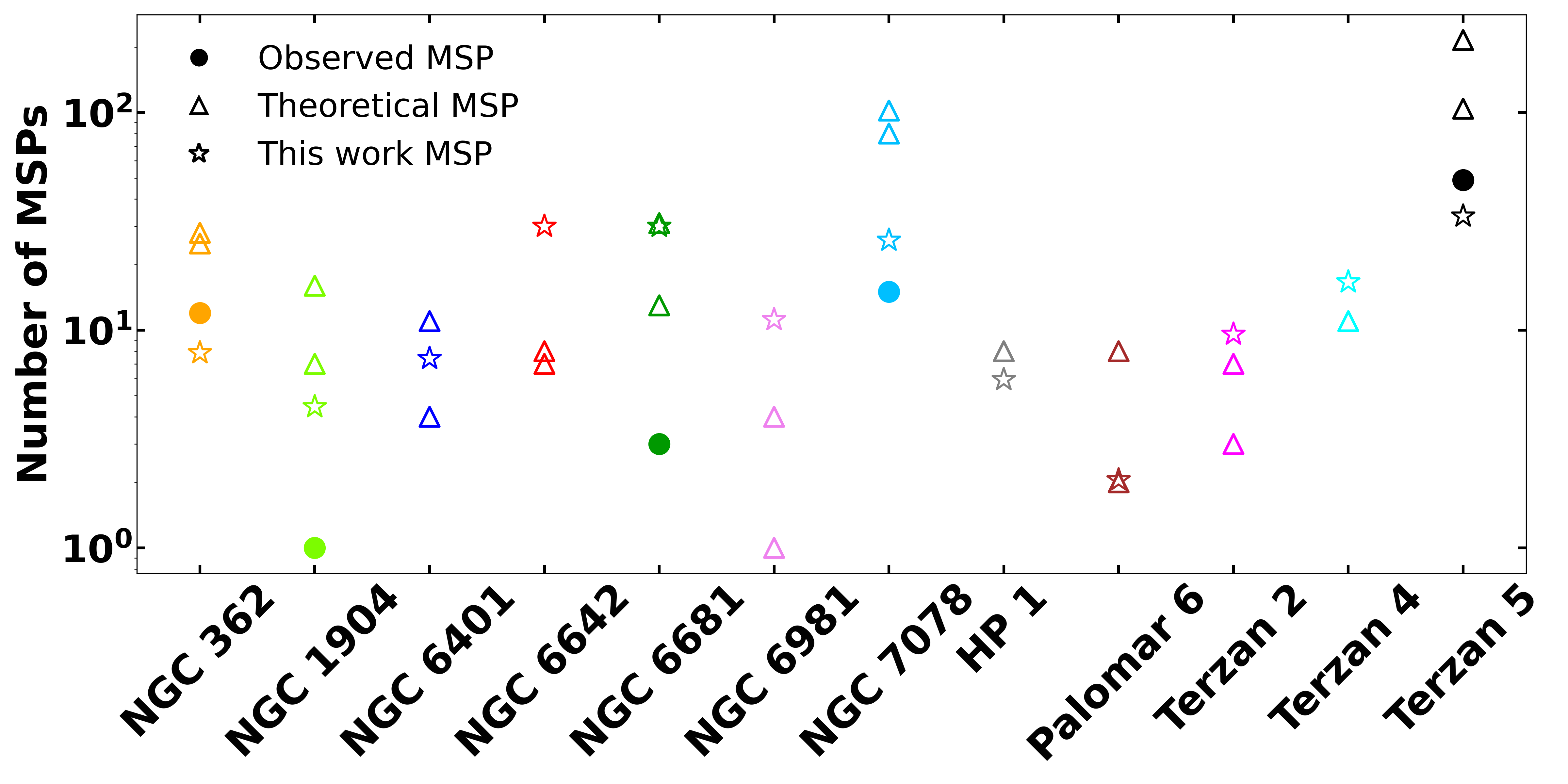}
\caption{ MSP populations in selected GCs. Top: MSP-to-bound NS ratio. Colours correspond to individual clusters as shown in the legend of Fig.~\ref{fig:dr-gc}, and colour dashed lines connect multiple estimates for the same cluster. Filled circles represent observed MSPs and open triangles show theoretical predictions. The red dashed line indicates the median value, while the shaded region represents the median absolute deviation (MAD) range. Lower panel: Number of MSPs in the selected GCs. Star symbols denote MSP numbers predicted in this work, computed as $N_{\mathrm{MSP}} = 0.0114 \times N_{\mathrm{Bound\ NS}}$ based on our simulation results.}
\label{fig:MSP-bound}
\end{figure}

In Fig.~\ref{fig:MSP-bound} (top), observed MSPs in real clusters are shown as coloured points, while theoretical MSP predictions ($N_{\rm MSP}$) based on a statistical model by \citet{Turk2013} and completed with FAST data \citep{Yin2024} are shown as open triangles. The corresponding numbers are listed in parentheses in Table~\ref{tab:ns-stat} (column~8). All displayed MSPs are gravitationally bound to their host clusters. For each cluster, we computed the ratio $N_{\rm MSP}/N_{\rm bound\,NS}$. These values are summarised in Table~\ref{tab:msp-ratios}. Here, multiple entries indicate distinct observational or predicted estimates. These values are directly correspond to those shown in Fig.~\ref{fig:MSP-bound} (top). With these ratios, we obtained a median MSP-per-NS coefficient of 
\begin{equation}  
k \equiv \mathrm{median}\!\left(N_{\rm MSP}/N_{\rm bound\,NS}\right) = 0.0114^{+0.0173}_{-0.0103}.
\end{equation}
Figure~\ref{fig:MSP-bound} (lower panel) shows the resulting MSP counts predicted for each cluster (with asterisks). To analyse the ratio of MSPs to bound NSs in selected GCs, we computed the MSP/NS ratio for both observed and theoretical data. The median and median absolute deviation (MAD) were employed instead of the mean and standard deviation, since these statistics are less sensitive to outliers and provide a more robust description of non-Gaussian datasets. 

The median represents the central value that divides the dataset into two equal halves, while MAD\ quantifies the typical deviation of data points from the median,  defined as $\mathrm{MAD} = \mathrm{median}\left(|x_i - \mathrm{median}(x)|\right).$

\begin{itemize}
    \item overall dataset (observed + theoretical): 
    $\mathrm{median} = 0.011449$, 
    $\mathrm{MAD} = 0.008142$;
    \item upper level of the predicted values for each cluster (\textit{theor2:} the greater of the two theoretical estimates per cluster, corresponding to predictions by \citet{Yin2024}; \textit{theor1} denotes predictions by \citet{Turk2013} and corresponding to the lower of the two  theoretically predicted values): 
    $\mathrm{median}_{\mathrm{theor2}} = 0.016154$, 
    $\mathrm{MAD}_{\mathrm{theor2}} = 0.012592$;
    \item observed values only: 
    $\mathrm{median}_{\mathrm{obs}} = 0.006608$, 
    $\mathrm{MAD}_{\mathrm{obs}} = 0.005467$.
\end{itemize}

Based on these results, the lower and upper bounds were defined as
$\mathrm{lower}=\mathrm{median}_{\mathrm{obs}}- \mathrm{MAD}_{\mathrm{obs}} = 0.001141$, $\mathrm{upper} = \mathrm{median}_{\mathrm{theor2}} + \mathrm{MAD}_{\mathrm{theor2}} = 0.028746.$
These values define the grey shaded region in the Fig.~\ref{fig:MSP-bound}, representing the statistically expected range of MSP/NS ratios. The red dashed line marks the overall median, while the grey dashed lines correspond to the upper and lower MAD-based limits. Observed MSP ratios are shown as filled circles, while theoretical predictions are shown as open triangles connected by dashed lines for each cluster.

We then applied the same coefficient to the NSs stripped from each system, assuming that the MSP-to-NS ratio is approximately conserved during the clusters’ dynamical evolution and subsequent deposition of NSs into the Galactic central bulge.

We extended the calibration to disrupted clusters by using the same coefficient, $k$, rather than cluster-specific values. Because the now-destroyed GCs lack reliable present-day parameters (e.g. mass, metallicity, encounter rate), a single empirically anchored \(k\) provides a consistent mapping from each hypothetical cluster’s neutron-star population to its expected MSP yield. Our simulations supply the full 6D phase-space distribution of deposited NSs, so the MSP-per-NS approach converts these kinematic populations to \(N_{\rm MSP}\) without invoking additional, poorly constrained global factors, yielding a robust and uniform estimate for the disrupted population.

\begin{table}[t]
\centering
\scriptsize
\caption{$N_{\rm MSP}/N_{\rm bound\,NS}$ ratios for each cluster used in the calibration for $\mathrm{obs/bound \ NS}$, $\mathrm{theor1/bound \ NS}$, and $\mathrm{theor2/bound \ NS}$.}
\label{tab:msp-ratios}
\resizebox{\columnwidth}{!}{%
\begin{tabular}{lccc}
\hline
\hline
Cluster & $\mathrm{obs/bound \ NS}$ & $\mathrm{theor1/bound \ NS}$ & $\mathrm{theor2/bound \ NS}$ \\
\hline
NGC~362   & 0.0174 & 0.0362 & 0.0406 \\
NGC~1904  & 0.0026 & 0.0179 & 0.0410 \\
NGC~6401  & --     & 0.0062 & 0.0169 \\
NGC~6642  & --     & 0.0027 & 0.0030 \\
NGC~6681  & 0.0011 & 0.0118 & 0.0049 \\
NGC~6981  & --     & 0.0010 & 0.0041 \\
NGC~7078  & 0.0066 & 0.0352 & 0.0449 \\
HP~1      & --     & 0.0000 & 0.0154 \\
Palomar~6 & --     & 0.0111 & 0.0444 \\
Terzan~2  & --     & 0.0036 & 0.0083 \\
Terzan~4  & --     & --     & 0.0075 \\
Terzan~5  & 0.0167 & 0.0355 & 0.0734 \\
\hline
\end{tabular}%
}
\end{table}

\subsection{Mock observations: NSs from real and destroyed GCs}\label{subsec:mock}

In Fig.~\ref{fig:XYZ-GCs-Real-Destr}, we show distribution of NSs in Cartesian Galactocentric coordinates ($X_{\star},\ Y_{\star},\ Z_{\star}$). To convert the individual NSs coordinates to Galactic coordinates (Heliocentric distance -- $d$, Galactic longitude (LON) -- $\ell$ , and latitude (LAT) -- $b$), we used Galactocentric coordinates of the Sun as $X_{\odot}= -8178 \, \text{pc}$ \citep{Gravity2019}, $Y_{\odot} = 0 \, \text{pc},\ Z_{\odot} = 20\, \text{pc}$ \citep{Bennett2019}. Next, we used Eq.~\ref{eq:heliocentric} to find the Cartesian Heliocentric coordinates of NSs ($x_{\star},\ y_{\star},\ z_{\star}$). After using Eq.~\ref{eq:galactic}, we transformed the NSs Heliocentric coordinates into Galactic coordinates. More details on a similar method are given in \cite{kalambay2022mock-mukha}, while \cite{Bissekenov-2024} used this method for simulated clusters from \cite{Shukirgaliyev2017, Shukirgaliyev2018,2019MNRAS.486.1045S, Shukirgaliyev2021}: 
\begin{equation}
\begin{gathered}
x_{\star} = X_{\star} - X_{\odot},\ \ \ \ \ \ \
y_{\star} = Y_{\star} - Y_{\odot}, \ \ \ \ \ \ \ 
z_{\star} = Z_{\star} - Z_{\odot} 
\end{gathered}
\label{eq:heliocentric}
,\end{equation}
\vspace{-1.5\baselineskip} 
\begin{equation}
\begin{gathered}
d = \sqrt{x_{\star}^2 + y_{\star}^2 + z_{\star}^2},\
\ell = \arctan\left(\frac{y_{\star}}{x_{\star}}\right),\  
b = \arcsin\left(\frac{z_{\star}}{d}\right)
\end{gathered}
\label{eq:galactic}
.\end{equation}

\begin{figure*}[htb!]
\centering
\includegraphics[width=0.95\linewidth]{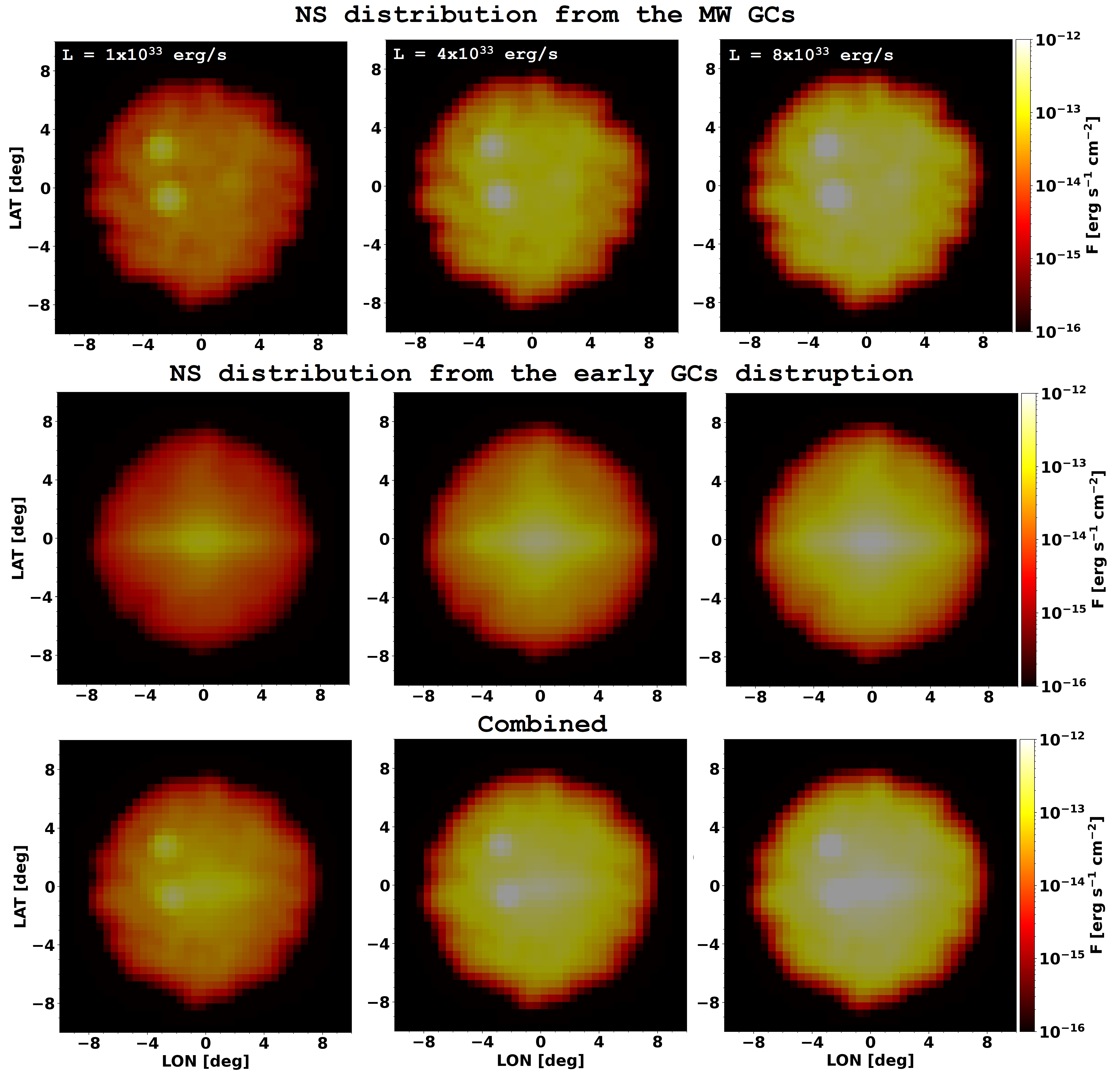}
\caption{$\gamma$-ray flux maps from simulated NSs within 1 kpc in Galactic coordinates. The flux is computed assuming a MSP luminosity of $\langle L\rangle \sim 1\times 10^{33}\, {\rm erg/s}$ (left), $\langle L\rangle \sim  4\times 10^{33}\, {\rm erg/s}$ (middle), and $\langle L\rangle \sim 8\times 10^{33}\, {\rm erg/s}$ (right), scaled by empirical factors for observed GCs (top), disrupted GCs (middle), and the combination of both populations (bottom). The colour scale indicates the total $\gamma$-ray flux in each bin, shown on a logarithmic scale between $10^{-12}$ and $10^{-16}\, {\rm erg\ s^{-1}\ cm^{-2}} $. To obtain a smoother flux map, we applied a Gaussian filter with a smoothing parameter of $\sigma=1$, which corresponds to a smoothing scale of approximately 0.5°.}
\label{fig:flux-MSP}
\end{figure*}

We combined the NS data from all our simulated clusters to obtain a distribution of NSs around the GalC in the Galactic coordinates. Then, the previously obtained coefficient ($k=0.0114$) was multiplied by the number of NSs in the centre of the Galaxy to estimate the number of MSPs within the 1 kpc sphere around the centre. MSPs in GCs have mean $\gamma$-ray luminosities roughly $\langle L\rangle \sim (1-8)\times 10^{33}\, {\rm erg/s}$ \citep{Amerio-2024}. For the purposes of demonstration, we used three values of luminosity $\langle L\rangle = 1\times 10^{33}\, {\rm erg/s}$, $\ 4\times 10^{33}\, {\rm erg/s}$, and $ 8\times 10^{33}\, {\rm erg/s}$ to estimate the total flux from the GalC. This result is shown in Fig.~\ref{fig:flux-MSP} (top). The same method was applied to NSs from destroyed clusters (Fig.~\ref{fig:flux-MSP}, middle), and the total flux (combination of NSs from both simulated real and destroyed theoretical GCs) is shown in the lower panel of Fig.~\ref{fig:flux-MSP}.

We want to highlight the interesting feature of the NS distribution from our population of the early destroyed theoretical GCs. The NS distribution from these clusters show a strong axisymmetric signature with the notable over-density in the Galactic disk plane, see Figs.~\ref{fig:XYZ-GCs-Real-Destr} and~\ref{fig:flux-MSP}. A similar spatial feature in the $\gamma$-ray excess was recently reported by  \cite{LV2025} and \cite{Muru2025}. Here, we also note another less prominent feature perpendicular to the Galactic plane; although this over-density is not as notable as the similar feature in Galactic disk, we like to call attention to it, as it might potentially be detected in future $\gamma$-ray observational programs and surveys. 

\begin{figure*}[htb!]
\centering
\includegraphics[width=0.99\linewidth]{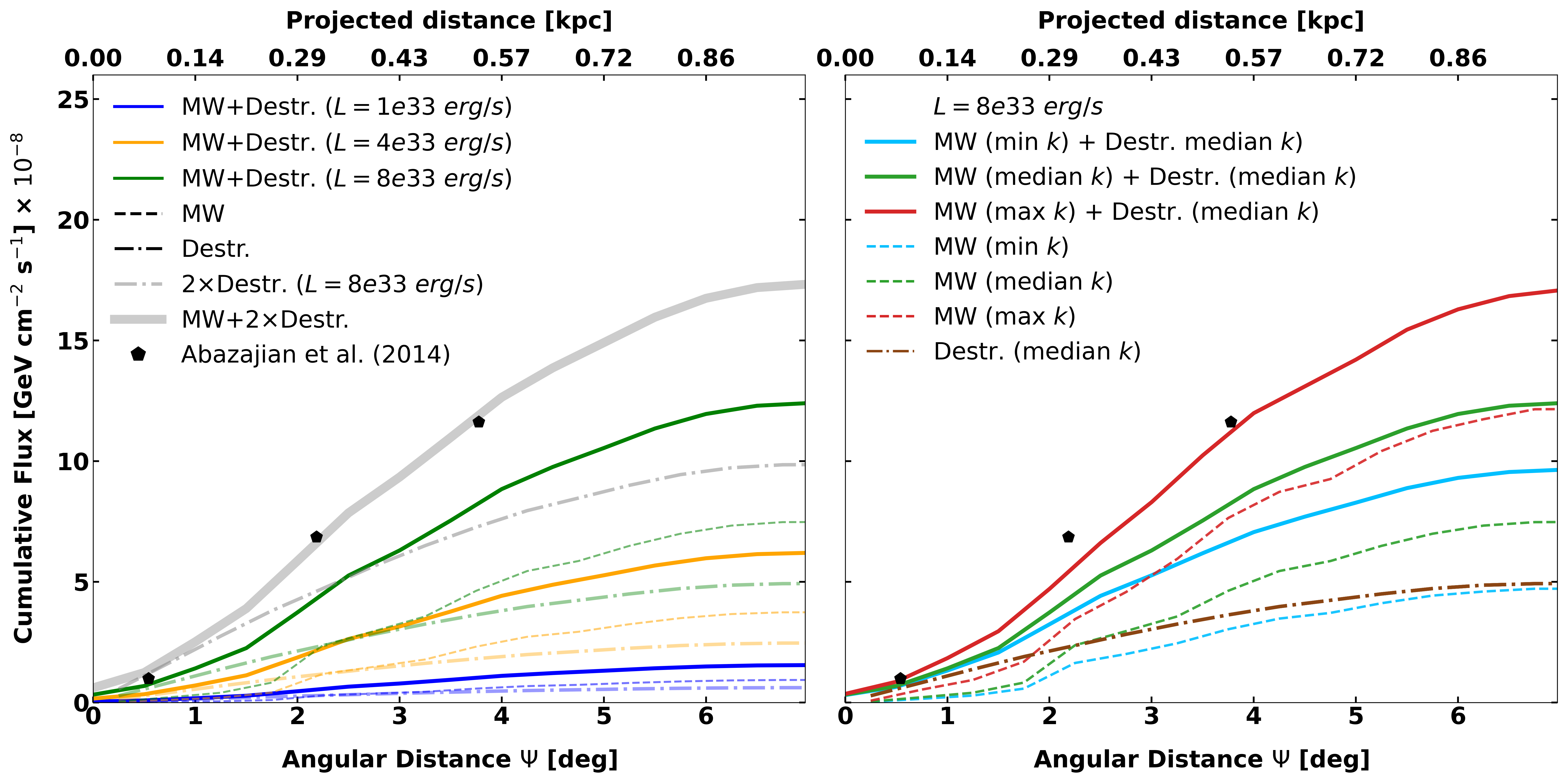}
\caption{Cumulative $\gamma$-ray flux as a function of projected radius from the GalC. The bottom axis shows the angular distance $\Psi$ in degrees, while the top axis indicates the corresponding projected distance in kiloparsecs, assuming a Sun–GalC distance of 8178 pc.
Left panel: Results for three different luminosities: $L = 1\times10^{33}\,\mathrm{erg\,s^{-1}}$ (blue), $L = 4\times10^{33}\,\mathrm{erg\,s^{-1}}$ (orange), and $L = 8\times10^{33}\,\mathrm{erg\,s^{-1}}$ (green).
Right: Case with fixed luminosity of $L = 8\times10^{33}\,\mathrm{erg\,s^{-1}}$, but different values of the scaling parameter $k$: the median value $k=0.0114$ (same green as on the left), the maximum individual value, corresponding to \textit{theor2} from \citet[shown in red]{Yin2024}, and the minimum individual value, corresponding to \textit{theor1} from \citet[shown in sky blue,]{Turk2013} as listed in Table~\ref{tab:msp-ratios} for each cluster. The legends in the figure: MW is the cumulative flux calculated only based on the selected real MW clusters (see Table~\ref{tab:init-param}); Destr.  refers to fluxes calculated only on the basis of the theoretical, already destroyed clusters; MW+Destr. is the sum of the flux from real MW plus the Destr. theoretical clusters; MW + 2 $\times$ Destr. is the same as the above sum, but with the contribution from the 'Destr.' clusters artificially enhanced by a factor of 2.}
\label{fig:cumsum-flux}
\end{figure*}

In Fig. \ref{fig:cumsum-flux}, we present the cumulative integrated $\gamma$-ray flux from the summed NSs in GalC. We summed up all the stripped NSs from the present-day observed GCs with the contribution of NSs from now-destroyed early GCs. We assumed a fixed luminosity of each MSP on three different levels: $\langle L\rangle =  1\times 10^{33}\, {\rm erg/s}$, $\ 4\times 10^{33}\, {\rm erg/s}$, and $ 8\times 10^{33}\, {\rm erg/s}$. In Fig. \ref{fig:cumsum-flux} (left), we plot the different contributions with different line-styles. Using different colours, we present the various luminosity levels. As we see from the result of our modelling, the summed contribution for the maximum luminosity level is generally quite consistent with the observationally available data \cite{Abazajian2014, Brandt2015}. In this plot, we also present the artificially enhanced by a contribution (of a factor of 2) from the early GC population (grey line:\ with MSP maximum luminosity level $8\times 10^{33}\, {\rm erg/s}$). In this case, we have nearly achieved an ideal correspondence with the observations after subtracting the contribution from the Galactic SMBH \citep{Abazajian2014}, represented by the filled black pentagons. 

In Fig. \ref{fig:cumsum-flux} (right panel), we present the same cumulative distribution, calculated using the minimum and maximum coefficients ({\it theor1} and {\it theor2}) from \citet{Turk2013} and \citet{Yin2024}, respectively. As can be seen, when adopting the maximum average flux $8\times 10^{33}\, {\rm erg/s}$ from a single MSP according to \cite{Amerio-2024} and using the maximum coefficient \textit{k} (red line), even with the current number of early (now already disrupted) GCs,  we are still able to account for the required overall level of the GCE.

For the lower luminosity levels, we see the significant deficit of our  theoretical cumulative flux compared with the observations. Such a flux deficit can mainly be explained by our applied assumptions: (i) on the total numbers of early GC; (ii) on the initial mass of these clusters; (iii) on the disruption rate of such systems (possibly accounting for intensive dynamical friction in initially more massive clusters). Additionally, in this work, we focussed on a scenario where MSPs were deposited into the GalC by GCs. However, pulsars could also form in situ, given the complex star formation history of the central region \citep{Schodel2020,Nogueras-Lara2021}. Moreover, \citet{Panamarev2019} showed that binaries in the inner few parsecs can harden efficiently (see their Fig.~10), potentially boosting MSP formation.

\section{Conclusions and discussion} \label{sec:disc-con}

We estimated the contribution to the $\gamma$-ray flux in the Galactic Centre from MSPs that originated in MW GCs. We used a suite of direct $N$-body simulations of observed GCs evolving in a time-varying MW potential to track the individual NSs tidally stripped from their host clusters and deposited in the central kpc region of the Galaxy. We also generated a set of hypothetical GCs that underwent complete tidal disruption and did not survive to the present day. We then estimated the number of MSPs per NS for both present-day and hypothetical GCs and used the mean $\gamma$-ray luminosity per pulsar to compute their net flux. In our numerical investigation, we found that:
\begin{itemize} 
    \item The total NS production scales with the initial GC mass (Fig.~\ref{fig:ratio-GCs-NS}, top), as expected for an applied common IMF. In contrast, the correlation between the bound NS and the current GC mass is weak (Fig.~\ref{fig:ratio-GCs-NS}, bottom), reflecting the fact that there is a significant kick velocity for the remnants during the formation of NS \citep{Banerjee2020, Kamlah2022MNRAS}. As a consequence, the slight dependence of the number of retained NSs on the cluster with the mass of the cluster. 
    \item 
    
    Calibrating the bound-NS counts in real clusters to their observed versus predicted MSP content yields the sample median coefficient $k=\rm{ median}( N_{\rm MSP}/N_{\rm bound\,NS})\simeq0.0114$ (Table~\ref{tab:msp-ratios}, Fig.~\ref{fig:MSP-bound}). Applying the same $k$ to all stripped NSs in the Galactic center provides a uniform mapping to $N_{\rm MSP}$ for both the sources of surviving and disrupted GC systems. The cumulative flux profile (Fig.~\ref{fig:cumsum-flux}, left panel) matches the observed trend when adopting higher MSP luminosities ($\langle L\rangle \sim 8\times10^{33}\,{\rm erg\,s^{-1}}$) and is brought into near-ideal agreement after subtracting the SMBH component \citep{Abazajian2014, Brandt2015} if the early-GC contribution is modestly enhanced (by factor of $\sim2$). For a more detailed analysis of the cumulative distribution of $\gamma$-ray flux around the GalC, we also employed two alternative ratios of the number of MSPs to the total number of NSs in observed Galactic stellar systems, namely, the coefficient $k$. One of them, {\it theor1} from \citet{Turk2013}, corresponds to the minimum value of this coefficient. The other, {\it theor2} from \citet{Yin2024}, corresponds to the maximum value of $k$. The results of these calculations are shown in Fig.~\ref{fig:cumsum-flux} (right panel). According to our estimates, even with the current number of early (i.e., now already disrupted) GCs, assuming the maximum {\it theor2} individual $k$ values for each observed GC, we were able to reproduce the overall level of $\gamma$--ray flux around the GalC at the observed level \citep{Abazajian2014, Brandt2015}.
    
    \item For lower $\langle L\rangle$ (e.g. $(1$ -- $4)\times10^{33}\,{\rm erg\,s^{-1}}$), the model under-produces the cumulative flux, pointing to uncertainties in the early GC populations, initial mass spectrum, phase-space distribution, disruption efficiencies, and dynamical friction. These are the parameters that directly modulate the stripped NS reservoir.

    \item Mock-observation maps within 1~kpc (Fig.~\ref{fig:flux-MSP}) show that the MSP population inferred from present-day GCs alone already contributes substantially to the GalC $\gamma$-ray signal.  Adding the disrupted population boosts both the amplitude and central concentration.

    \item  NSs deposited by early destroyed GCs exhibit an axisymmetric signature with a pronounced over-density along the disk plane (Figs.~\ref{fig:XYZ-GCs-Real-Destr} and \ref{fig:flux-MSP}) are qualitatively consistent with recent reports of a similar structure in the $\gamma$-ray excess morphology \cite{LV2025, Muru2025}. We also note another less prominent feature perpendicular to the Galactic plane, which can potentially be detected in future $\gamma$-ray observational surveys.
\end{itemize}

Thus, our modelling favours the MSP origin of the GCE over the DM-annihilation scenario. 
To estimate the net flux from pulsars, we assumed a mean $\gamma$-ray luminosity of $\langle L\rangle = 8\times 10^{33}\,{\rm erg\,s^{-1}}$. This assumption is critical for the total flux calculation. For example, \citet{Fragione2018} found that the observed GCE flux can be reproduced using a more moderate mean luminosity of $\langle L\rangle = 2\times 10^{33}\,{\rm erg\,s^{-1}}$. The main difference between our calculations and those performed by \citet{Fragione2018} is the more self-consistent treatment we adopted with respect to the number of NSs retained in the Galactic Centre region (<1 kpc). In that work, the authors assumed a relatively large and fixed retained fraction of $\sim10\%$ of the total number of produced NSs. In our investigation, we modelled each NS orbit individually and dynamically compute the retained fraction for each cluster. In this way, we found that only $\sim6\%$ of NSs remain within the central 1~kpc region of the Galaxy. Additionally, including a larger sample of potentially destroyed clusters could increase the number of deposited NSs in the central region, thereby relaxing the requirement for a relatively high average per-pulsar luminosity to match the observed flux. Finally, the MSP population in the Galactic Centre might also be supplemented by in situ formation, given the complex star formation history of the region \citep{Schodel2020,Nogueras-Lara2021} and efficient dynamical hardening of binaries in the inner few parsecs could alsio promote MSP production \citep[e.g.,][]{Panamarev2019}. A future study that jointly accounts for these effects could help mitigate the associated uncertainties.

The following extensions will make our predictions more precise. First, incorporating a fully time-varying Galactic potential directly in a code which features accurate treatment of binary stars, such as \textsc{NBODY6++GPU} \citep{Wang2015, Wang2016, Panamarev2019, Kamlah2022}, together with binary stellar evolution (to follow LMXB stages, recycling, and MSP birth+aging) would remove our reliance on a global $k$ and enable cluster-by-cluster predictions of $N_{\rm MSP}$ and their $\gamma$-ray luminosities. Second, implementing dynamical friction self-consistently (or coupling it to an evolving live background) should tighten the disrupted-GC contribution and central concentration. Third, we assumed a coefficient $k$ and also fixed the per-pulsar average luminosity $\langle L\rangle$. A hierarchical treatment that fits a cluster-dependent $k$, assuming individual natal-kick retention, and a full MSP luminosity function would reduce systematics tied to selection effects in current MSP/LMXB catalogues. Finally, forward-modelling MSP beaming geometry, duty cycles, and magnetospheric cut-offs would further reduce biases in the flux maps and in the cumulative profile. 

Despite these modelling choices, in its current state, our work favours an MSP origin of the GalC $\gamma$-ray excess over DM annihilation. This is primarily because the combined contribution of MSPs delivered by surviving and disrupted GCs naturally reproduces both the amplitude and concentration of the observed signal under reasonable assumptions. 

Further observational leverage will come from deep bulge radio pulsar searches (e.g. FAST, MeerKAT, SKA) targeted along our predicted over-densities \citep{Han2021GPPSI,Han2024GPPSVI,Frail2024MeerKAT,Keane2015SKA}; TeV-halo mapping with CTA to test the integrated MSP population \citep{Hooper2018TeVMSP}; improved MeV--GeV spectroscopy and diffuse-background control with next-generation $\gamma$-ray missions \citep{Caputo2022AMEGOX,DeAngelis2017eASTROGAM}; updated LMXB inventories (e.g. eROSITA/Chandra/NuSTAR) to anchor progenitors \citep{Jonker2011GBS,Wevers2016GBSopt,Hong2016NuSTARGC}; continuous-GW searches constraining the bulge MSP population \citep{Miller2023cGW}; and refined bulge and bar stellar maps to compare with our predicted morphology \citep{Wegg2013VVV,Valenti2016VVV}. These data will allow for tighter, physically anchored priors to be set on MSP demographics, thereby improving our understanding of the Galactic $\gamma$-ray excess content.

\begin{acknowledgements}

The authors thank the anonymous referee for a very constructive report and suggestions that helped significantly improve the quality of the manuscript. This research has been funded by the Science Committee of the Ministry of Science and Higher Education, Republic of Kazakhstan (Grant No. AP26102895). PB and MI appreciate the support from the Polish Academy of Sciences Special Program and the U.S. National Academy of Sciences Special Program under the Long-term program to support Ukrainian research teams grant No.~PAN.BFB.S.BWZ.329.022.2023. We also gratefully acknowledge the Polish high-performance computing infrastructure 
PLGrid (HPC Center: ACK Cyfronet AGH) for providing computer facilities and 
support within computational grant No.~ PLG/2026/019243

\end{acknowledgements}

\bibliographystyle{bibtex/aa}  
\bibliography{bibtex/sources.bib}   

\end{document}